# Field dependent neutron diffraction study in $Ni_{50}Mn_{38}Sb_{12}$ Heusler alloy

Roshnee Sahoo[1,2*], Amitabh Das[3], Norbert Stuesser[4], K. G. Suresh[1]

[1]Department of Physics, Indian Institute of Technology Bombay, Mumbai 400076, India

[2]Max Planck Institute for Chemical Physics of Solids, Nöthnitzer Str. 40, 01187 Dresden, Germany

[3]Solid State Physics Division, Bhabha Atomic Research Centre, Mumbai 400085, India

[4]Helmholtz-Zentrum Berlin, Hahn-Meitner-Platz 1, D-14109 Berlin, Germany

**Abstract**:

In this paper, we present temperature and field dependent neutron diffraction (ND) study to unravel the structural and the magnetic properties in $Ni_{50}Mn_{38}Sb_{12}$ Heusler system. This alloy shows martensitic transition from high temperature austenite cubic phase to low temperature martensite orthorhombic phase on cooling. At 3 K, the lattice parameters and magnetic moments are found to be almost insensitive to field. Just below the martensitic transition temperature, the martensite phase fraction is found to be 85%. Upon applying the field, the austenite phase becomes dominant, and the field induced reverse martensitic transition is clearly observed in the ND data. Therefore, the present study gives an estimate of the strength of the martensite phase or the sharpness of the martensitic transition. Variation of individual moments and the change in the phase fraction obtained from the analysis of the ND data vividly show the change in the magneto-structural state of the material across the transition.

*Corresponding author: Dr. Roshnee Sahoo (e-mail: sahoo@cpfs.mpg.de).

**Introduction:**

Recently, there has been a lot of interest in $Ni_{50}Mn_{50-x}Sb_x$ based full Heusler system showing to their various functional properties such as magnetocaloric effect (MCE), shape memory effect (SME) and magnetoresistance (MR).[1-5] Martensitic transition is the key factor for such exciting properties. Inoff-stoichiometric $Ni_{50}Mn_{50-x}Sb_x$, the martensitic transition is observed in narrow composition range for $x$ = 12, 13 and 14.[4] In these compositions, the excess Mn occupies the Sb site(4b site), leading to an antiparallel alignment of moments between the Mn at 4a site and Mn at 4b site.[6] The martensitic transition is followed by a change in the magnetic state; this magneto-structural effect is the underlying reason for the functional properties. The martensitic transition and the multifunctional properties can be controlled by external parameters like temperature, pressure and magnetic field. After, the discovery of magnetic field induced shape memory effect in NiMnIn system,[7] there have been many efforts to induce martensitic transition by applying a magnetic field. The effect of magnetic field on the martensitic transition has been studied in many NiMn based systems such as NiCoMnAl, NiMnSn and NiCoMnSb.[8-10] These systems can be very promising for magnetically controlled devices such as sensors, actuators and magnetic refrigerators.[11-14] Therefore, exploring such materials is important from the point of view of fundamental aspects as well as application potential. In this regard NiMnSb system can also be a promising candidate. The room temperature magneto-structural transition and the low cost of the constituent elements make this system more attractive for room temperature applications.

Neutron diffraction(ND)is an important probe for studying crystal as well as magnetic structures, more so in the case of systems with strong magnetostructural coupling. In such systems, ND can provide information, which can otherwise be obtained by magnetic and structural studies. Recently, we have reported ND experiments in $Ni_{50}Mn_{38}Sb_{12}$ and $Ni_{45}Co_5Mn_{38}Sb_{12}$ Heusler systems.[6] In these alloys, we found that the high temperature crystallographic phase exhibits cubic $L2_1$ structure in which Mn moments are aligned ferromagnetically. Below, the martensitic transition temperature, the phase is of orthorhombic Pmma structure. A similar behaviour has also been reported in $Ni_2Mn_{1.48}Sb_{0.52}$ system by Brown et al.[15] In many NiMn based Heusler alloy, field induced irreversibility is reported across the first order martensitic transition. This field driven austenite phase gets frozen, which prohibits the transformation of parent phase to the martensite phase.[16-18] In a previous study, field induced irreversibility was reported by Nayak et al in NiCoMnSb system using resistivity, heat capacity and magnetization technique.[10] At temperatures near martensitic transition, it has been established that field induced transformation from martensite to austenite phase occurs and the transition is irreversible. However, ND is a unique probe that has not been utilized so far to verify this irreversibility. To shed more light, on the irreversibility and to understand the effect of magnetic field on the first order martensitic phase transition and related effects, we present the results of in-field neutron diffraction studies in $Ni_{50}Mn_{38}Sb_{12}$ alloy.

**Experimental Details:**

Polycrystalline $Ni_{50}Mn_{38}Sb_{12}$ ingot of 5 g was prepared by arc melting process under argon atmosphere using Ni, Mn and Sb of atleast 99.99% purity. The ingot was melted several times for homogeneity. Extra Mn was taken to compensate the weight loss during arc melting. The sample was subsequently annealed in evacuated quartz tube at 850˚C for 24 hr for homogenization. The sample composition was verified from EDAX analysis. The magnetization measurements were carried out using a vibrating sample magnetometer attached to a Physical Property Measurement System (Quantum Design, PPMS-6500). For the temperature variation of magnetization, the measurement was done at the rate of 4K/min. For field dependence of magnetization, the sample was initially cooled down to the lowest possible temperature in zero field and then heated back to the desired measuring temperature and the measurement was done with a field ramp of 150 Oe. In order to carry out the neutron measurement, the bulk sample was crushed into powder and reannealed to relieve lattice strain or dislocation. The magnetization measurements were also performed on powder.The neutron diffraction study was carried out using on the E6 neutron diffractometer (λ=2.4 A°) at HZB in Berlin, Germany. The powdered sample was kept in anAluminium can, which was placed in the orange-type cryostat.The temperature was varied in the range of 3-340 K. The neutron diffraction pattern refinement was been carried out using Fullprof program.[19] In all magnetic refinement only the Mn occupied site moment has been considered for refinement. In order to minimize the error,the refinement of the excess Mn occupied at Sb site is avoided.

**Results and Discussion:**

At low temperatures,the martensite phase exhibits orthorhombic structure (Pmma space group) shown in figure 1(a).The unit cell as shown in here is for stoichiometry composition $Ni_{50}Mn_{25}Sb_{25}$. In this unit cell, Ni occupies the 4h *(0, y, ½)* and 4k *(¼, y, z)* Wyckoff site. 2a*(0, 0, 0)* and 2f*(¼, ½, z)* sites are occupied by Mn atoms. Sb atom occupies 2b *(0, ½, 0)* and 2e *(¼, 0, z)* sites. In the off stoichiometry, the excess Mn atoms also occupy Sb site. On the other hand, the high temperature austenite phase is cubic (Fm3m space group),in which 4a *(0, 0, 0)* site is occupied by Mn atoms and 4b *(½, ½, ½)* site is occupied by Sb and 8c *(¼, ¼, ¼)* site is occupied by Ni. Similar to the orthorhombic phase, the cubic phase with off stoichiometry shows partial occupancy of Mn in Sb site. The detailed structural analysis is reported in our previous work.[6]

Figure 1(b) shows the temperature variation of magnetization under zero field cooled (ZFC) and field cooled cooling (FCC) mode in field of 0.05 T. In ZFC mode, the sample was first cooled down to the lowest temperature in zero field and the magnetization data was taken during heating. Subsequently, in the FCC mode, the data was collected while cooling down in the same field. For both ZFC and FCC curves, the data was taken at a rate of 4 K/min. In ZFC curves, the transition temperatures are denoted as $A_S$ (austenite start) and $A_F$ (austenite finish). Similarly, $M_S$ (martensite start) and $M_F$ (martensite finish) are indicated in FCC curve. Figure 1(c) depicts the magnetic isotherms at 3 K and near transition region (such as 320, 330 and 340 K). It can be observed that the magnetization saturates at 3 K with a value of 1.5 $\mu_B$/f.u..

At 3 K, the ND pattern of $Ni_{50}Mn_{38}Sb_{12}$ is shown for different fields (figure 2a). These patterns are taken for 0, 5 T and again for zero field. Lattice parameters and magnetic moment of martensitic phase are obtained by profile matching from Fullprof program. We do not observe any noticeable variation in the lattice parameters (within the resolution of the instrument). But, there is slight increase in Mn site moment on application of field. It can be noticed that there is no signature of austenite phase even for the highest field. This result supports the fact that 5 T may not be enough to obtain the reverse martensitic transition at low temperatures. A similar result is reported in NiMn based Heusler alloys, where it was found that the field induced irreversibility at 4.2 K requires very high field for the reverse martensitic transition.[20]

**Table 1.** Refined structural parameters from analysis of neutron diffraction data on $Ni_{50}Mn_{38}Sb_{12}$ at 3 K and 320 K. FM moment per Mn atom at 3 K.

| T = 3 K | Orthorhombic (Pmma) | | | | | Moment ($\mu_B$) | | |
|---|---|---|---|---|---|---|---|---|
| | a (Å) | b (Å) | c (Å) | $R_B$ | $R_F$ | 2a | 2f | Rm |
| $\mu_0H$=0 T | 8.429(1) | 5.452(1) | 4.299(1) | 8.3 | 6.2 | 3.0(3) | 2.5(3) | 9 |
| $\mu_0H$=5 T | 8.428(1) | 5.451(1) | 4.299(1) | 8.2 | 6.1 | 3.7(3) | 2.9(2) | 10 |
| $\mu_0H$=0 T | 8.429(1) | 5.451(1) | 4.298(1) | 6.9 | 7 | - | - | - |

| T = 320 K | Orthorhombic (Pmma) | | | | | Cubic (Fm3m) | | |
|---|---|---|---|---|---|---|---|---|
| | a (Å) | b (Å) | c (Å) | $R_B$ | $R_F$ | a (Å) | $R_B$ | $R_F$ |
| $\mu_0H$=0 T | 8.418(2) | 5.563(1) | 4.257(1) | 8.6 | 8.1 | - | - | - |
| $\mu_0H$=2T | 8.414(1) | 5.568(1) | 4.254(1) | 10.2 | 8.9 | 5.978(3) | 5.5 | 7.7 |
| $\mu_0H$=5T | 8.419(2) | 5.564(1) | 4.259(1) | 11.4 | 9 | 5.959(5) | 5.7 | 7.6 |
| $\mu_0H$=0 T | 8.416(1) | 5.568(1) | 4.255(1) | 9.3 | 8.2 | 5.984(5) | 5 | 6.3 |

*$R_B$ and $R_F$ are structural reliability factor.

*$R_M$ is magnetic reliability factor.

Figure 2b shows the ND pattern at 320 K in zero field. The inset of the figure shows the variation of a selected region of the pattern in different fields. The data is also taken after removal of the field. The temperature of 320 K signifies the austenite start temperature and at this temperature the diffraction pattern is indicative of orthorhombic martensite structure. The structural parameters are given in table 1. It can be seen that for 2T the intensity of (201) martensite reflection increases and in addition a weak reflection appears at 48°, which can be seen in the inset of figure 2(b). This indicates the emergence of the austenite (200) peak, which is quite clear at 5 T. Also, the intensity of the martensite (201) reflection is lower in 5 T, as compared to that in 2 T. The initial increase in intensity of martensite reflection (at 2 T) maybe due to the martensite reorientation. And the later decrease (at 5 T) is due to the growth of austenite phase. A similar observation is also reported in $Ni_{49.7}Mn_{34.3}In_{16}$ Heusler alloy.[21]

As given in table 1, at 320 K also, the lattice parameters of orthorhombic phase donot show much change with application of field. For 5 T, a fraction of austenite phase emerges in the martensite regime, whose structural parameters are given in table 1. As this temperature is just close to the austenite start temperature, the austenite phase moment value is negligible for zero field and even for 2 T. In 5 T, the austenite phase moment has appeared 0.7(1) $\mu_B$ for the 4a site.

Figure 3 shows the ND patterns for 330 K in different fields namely 0, 2 T, 5 T and again inzero field. Results of structural refinement with field are shown in table 2. It can be seen that, in the martensite phase the lattice parameters vary with field; however those in the austenite phase remain unchanged.

This selected temperature (330 K) is very well into the martensitic transition region. It can be seen that inzero field there is a weak (200) austenite peak along with the martensite (201) peak. With increase in field, the intensity of this austenite peak increases, while that of the martensite peak decreases. At this temperature, the field induced reverse transition from martensitic (orthorhombic) to austenite (cubic) phase is clearly visible. From the structural refinement, the martensite and austenite fractions as a function of field are calculated and are given in figure 4. It is clear that with increase in field, the austenite phase fraction increases. For zerofield, the austenite fraction is 15% which increases to 70% on application of 5T. It has been observed that the austenite fraction remains the same even after making the field zero. This demonstrates the field induced irreversibility. A similar result is also reported in NiCoMnSb system, which shows irreversibility in magnetization, transport and heat capacity isotherms.[10]

**Table 2.** Structural parameters at T=330 K for different field values from neutron diffraction refinement for $Ni_{50}Mn_{38}Sb_{12}$ system

| T = 330 K | Orthorhombic (Pmma) | | | | | | Cubic (Fm3m) | | | |
|---|---|---|---|---|---|---|---|---|---|---|
| | a (Å) | b (Å) | c (Å) | $R_B$ | $R_F$ | M(%) | a (Å) | $R_B$ | $R_F$ | A(%) |
| H=0 T | 8.421(2) | 5.567(1) | 4.256(1) | 12.3 | 11.2 | 85 | 5.854(1) | 6.1 | 5.6 | 15 |
| H=2T | 8.424(3) | 5.564(2) | 4.254(1) | 15.7 | 13.1 | 60 | 5.850(1) | 6.8 | 4.6 | 40 |
| H=5T | 8.418(4) | 5.564(2) | 4.248(2) | 22.2 | 20 | 30 | 5.850(1) | 5.3 | 4.2 | 70 |
| H=0 T | 8.402(4) | 5.562(2) | 4.265(2) | 15.9 | 12 | 30 | 5.849(1) | 3.4 | 2.8 | 70 |

M(%) and A(%) signifies the fractional percentage of martensite or austenite phase.

From the lattice parameters of both austenite and martensite phase, $\lambda_2$ parameter can be computed from Bain transformation matrix (U).[22] Here, $\lambda_2$ is the middle eigenvalue of the matrix. There is significant relation of $\lambda_2$ and hysteresis. If $\lambda_2=1$, it suggests unstressed and untwinned interface between the two phases. And when $\lambda_2 \neq 1$, the growth of austenite is accompanied by finely twinned martensite. In NiCoMnSn based Heusler alloy, structural transformation from monoclinic to cubic phase is occurred. In this system $\lambda_2$ value is reported to be 1.[23] Here, at 330 K solving Bain transformation matrix, $\lambda_2$ the middle eigenvalue has been calculated as 1. This result suggests an infinite number of compatible interfaces between austenite and finely twinned martensite phases, and the thermal hysteresis is expected to be lower. Usually, system with weak first order transition often leads to relatively low hysteresis. But the hysteresis also depends on various other factors such as other eigenvalues ($\lambda_1$, $\lambda_3$), interfacial energy, transition temperature and the latent heat.

**Table 3.** Site specific magnetic moment values at T=330 K and 340 K for different fields obtained from the neutron diffraction data refinement in $Ni_{50}Mn_{38}Sb_{12}$

| T=330K | Moment($\mu_B$)(Pmma) | | | Moment($\mu_B$)(Fm3m) | | T=340K | Moment($\mu_B$)(Fm3m) | |
|---|---|---|---|---|---|---|---|---|
| | 2a | 2f | Rm | 4a | Rm | | 4a | Rm |
| $\mu_0H=0T$ | 0.7(7) | 0.7(7) | 11 | 0.8(1) | 12 | $\mu_0H=0T$ | 0.4(1) | 11 |
| $\mu_0H=2T$ | 0.7(5) | 1.9(5) | 25 | 0.9(1) | 11 | $\mu_0H=2T$ | 0.8(1) | 15 |
| $\mu_0H=5T$ | - | - | - | 0.7(1) | 6 | $\mu_0H=5T$ | 0.8(1) | 6 |

From the magnetic refinement, the martensite and austenite phase moments for 330 K are obtained and are shown in table 3. It can be observed that 2a and 2f site moments corresponding to the martensite phase increases with applied field. The moment values of 2a and 2f site are 0.7(7) $\mu_B$ and 0.7(7) $\mu_B$ in zero field. As these error values are large, these moment values are negligible. It has been observed that with increase in field, these moments vary to 0.7(5) $\mu_B$ and 1.9(5) $\mu_B$. The moment at 2f site increases with field of 2 T. However, for the austenite phase, it has been observed that moments corresponding to 4a sites remain constant. This is expected as at this temperature, the austenite phase is ferromagnetic. At 340 K near the Curie temperature of the austenite phase, upon applying field of 2 T, an increase in moment from 0.4(1) to 0.8(1)$\mu_B$ is observed and is shown in table 3.

Therefore, it can be seen that the temperature and field dependent ND study gives an estimate of the strength of the martensite phase or the sharpness of the martensitic transition. Variation of individual site moments and the change in the phase fraction obtained from the analysis of the ND data vividly show the change in the magneto-structural state of the material across the transition.

**Conclusion:**

In summary, field dependent neutron diffraction study has vividly revealed the effects of supercooling/heating and the field induced irreversibility associated with the first order martensitic transition in $Ni_{50}Mn_{38}Sb_{12}$ alloy. This has been done by comparing the diffractograms at 3, 320 and 330 K. It is found that when the field is applied at temperatures much below the transition temperature, a field of 5T is not sufficient to induce the reverse martensitic transition. A stable martensitic phase is observed at 3K, where the lattice parameter value remains unchanged with the applied field. On the other hand at 330 K, the reverse martensitic transition is clearly observed with 5T field. The middle eigenvalue $\lambda_2$ has been calculated as 1. The evolution of the austenite phase at the cost of the martensite phase with increase in field has been elucidated.

**References:**

[1]Y. Sutou, Y. Imano, N. Koeda, T. Omori, R. Kainuma, K. Ishida & K. Oikawa, Appl. Phys. Lett. **85,** 19 (2004).

[2]M. Khan, I. Dubenko, S. Stadler and N. Ali, J. Appl. Phys. **101,** 053919 (2007).

[3]A. K. Nayak, K. G. Suresh and A. K. Nigam, J. Phys. D: Appl. Phys. **42**,035009 (2009).

[4]J. Du, Q. Zheng, W. J. Ren, W. J. Feng, X. G. Liu and Z D Zhang, J. Phys. D: Appl. Phys. **40**, 5523 (2007).

[5]R. Sahoo, A. K. Nayak, K. G. Suresh and A. K. Nigam, J. Appl. Phys. **109**, 123904 (2011).

[6]R. Sahoo, K. G. Suresh and A. Das, J. Mag. & Mag. Mat.**371**, 94 (2014).

[7]R. Kainuma, Y. Imano, W. Ito, Y. Sutou, H. Morito, S. Okamoto, O. Kitakami, K. Oikawa, A. Fujita, T. Kanomata and K. Ishida, Nature. **439**, 957 (2006)

[8]R. Kainuma, W. Ito, R. Y. Umetsu, K. Oikawa, K. Ishida, Appl. Phys. Lett. **93**, 091906 (2008).

[9]K. Koyama, K. Watanabe, T. Kanomata, R. Kainuma, K. Oikawa and K. Ishida, Appl. Phys. Lett. **88**, 132505 (2006).

[10]A. K. Nayak, K. G. Suresh and A. K. Nigam, Appl. Phys. Lett.**96**, 112503 (2010).

[11]A. Sozinov, A. A. Likhachev, N. Lanska and K. Ullako, Appl. Phys. Lett.**80**, 1746 (2002).

[12]S. J. Murray, M. A. Marrioni, A. M. Kukla, J. Robinson, R. C. O' Handley and S. M. Allen, J. Appl. Phys. **87**, 5774 (2000).

[13]H. E. Karaca, I. Karaman, B. Basaran, Y. Ren, Y. I. Chumlyakov, H. J. Maier. Adv Funct Mater. **19**, 983 (2009).

[14]G. H. Yu, Y. L. Xu, Z. H. Liu, H. M. Qiu, Z. Y. Zhu, X. P. Huang,L. Q. Pan, Rare Metals.**34**, 527 (2015).

[15]P. J. Brown, A. P. Gandy, K. Ishida, W. Ito, R. Kainuma, T. Kanomata, K. U. Neumann, K. Oikawa, B. Ouladdiaf, A. Sheikh and K. R. A. Ziebeck, J. Phys.: Condens. Matter. **22,** 096002 (2010).

[16]A. K. Nayak, M. Nicklas,C. Shekhar and C. Felser, 113, 17E308 (2013).

[17]A. K. Nayak, M. Nicklas, S. Chadov, C. Shekhar, Y. Skourski, J. Winterlik and C.Felser, Phys. Rev. lett **110**, 127204 (2013).

[18]V. K. Sharma, M. K. Chattopadhyay and S. B. Roy, Phys. Rev. B. 76, 140401 (2007).

[19]J. Rodriguez-Carvajal, Physica B. **192**, 55 (1993).

[20]A. K. Nayak, C. Salazar Mejia, S. W. D'Souza, S. Chadov,Y. Skourski, C. Felser and M. Nicklas, Phys. Rev. B. **90**, 220408(R) (2014).

[21]T. Krenke, E. Duman, M. Acet, E. F. Wassermann, X. Moya, L. Manosa, A. Planes, E. Suard and B. Ouladdiaf, Phys. Rev. B. **75**, 104414 (2007).

[22]Z. Zhang, R. D. James and S. Müller, Acta Mater. **57**, 4332 (2009).

[23]V. Srivastava, X. Chen and R. D. James, Appl. Phys. Lett. **97**, 014101 (2010).

**Figure Captions:**

FIG. 1. (a) Unit cell of martensite and austenite phases of $Ni_{50}Mn_{25}Sb_{25}$. (b) Temperature dependence of magnetization for $Ni_{50}Mn_{38}Sb_{12}$ in 0.05 T. (c) Magnetic isotherms for different temperatures for $Ni_{50}Mn_{38}Sb_{12}$. Theinset shows the magnetization curve for 330 K.

FIG. 2. Neutron diffraction pattern for $Ni_{50}Mn_{38}Sb_{12}$ at (a) 3 K and (b) 320 K (Inset shows the variation of martensite (201) reflection for $\mu_0 H$= 0, 2, 5 and back to 0 T).

FIG. 3. Neutron diffraction patterns in different fields for $Ni_{50}Mn_{38}Sb_{12}$ at 330 K in 0,2 T, 5 T and back to zero field.

FIG. 4. Martensite and austenite phase fractions (%) in $Ni_{50}Mn_{38}Sb_{12}$ alloy at 330 K.

Fig. 1

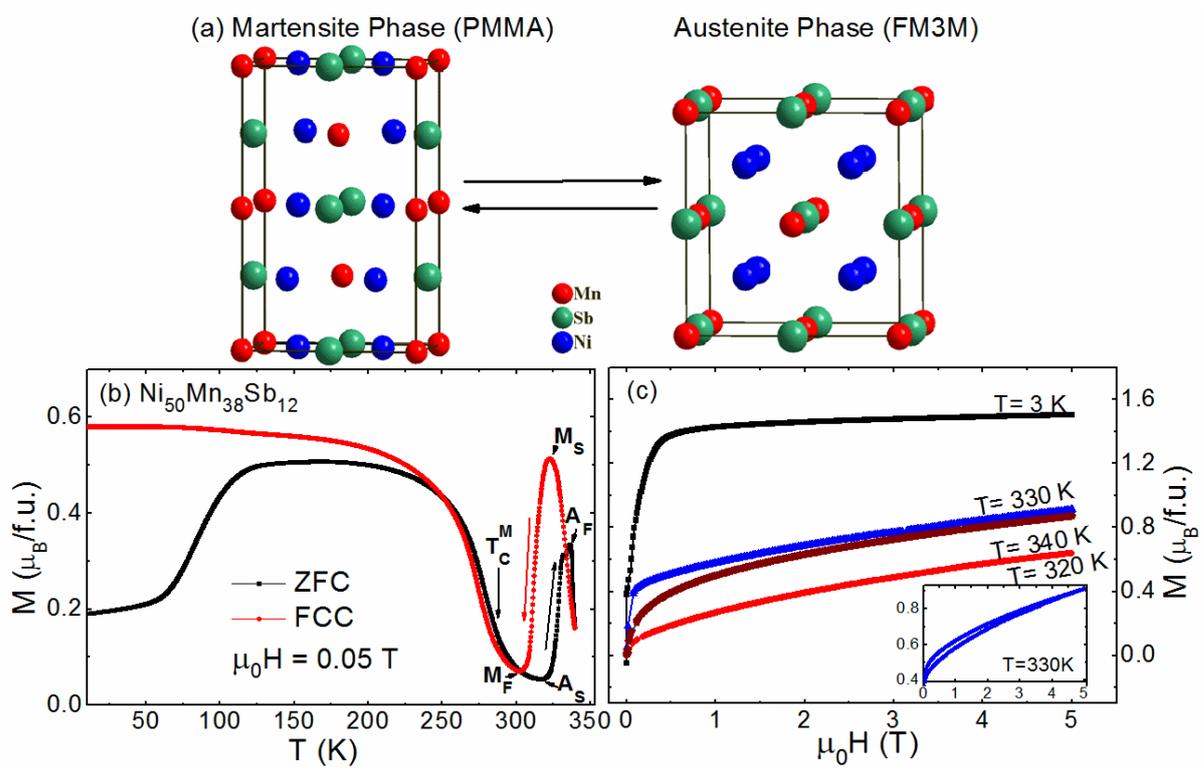

Fig. 2

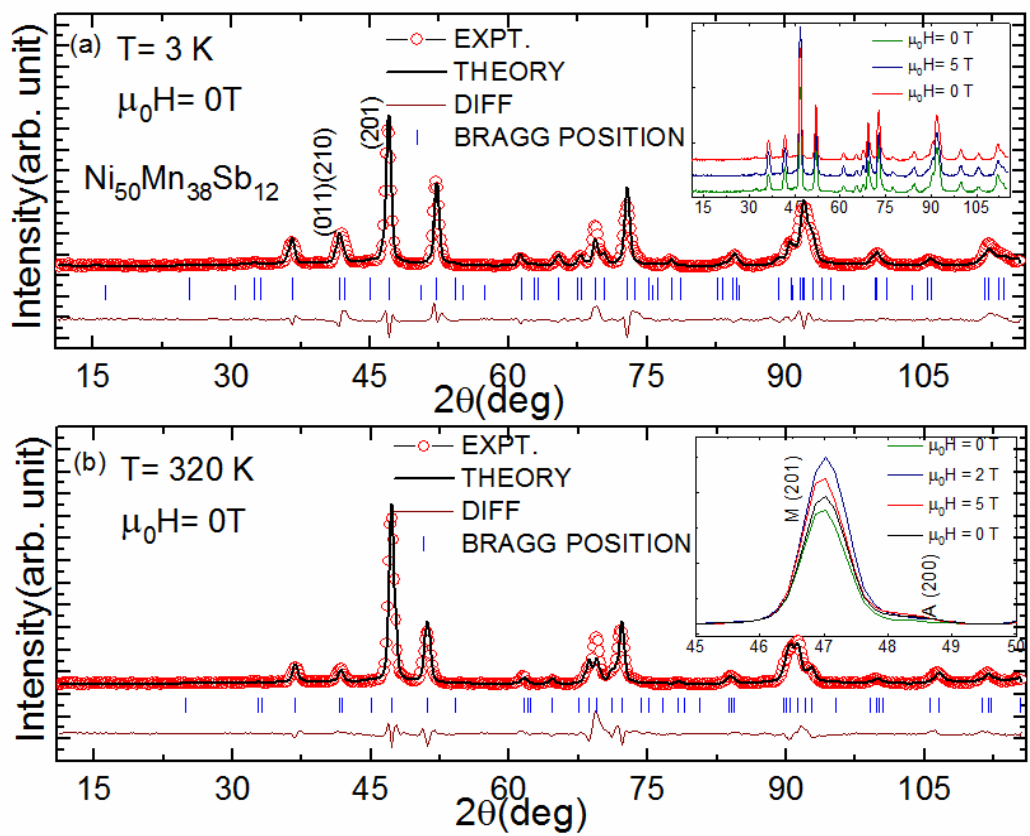

Fig. 3

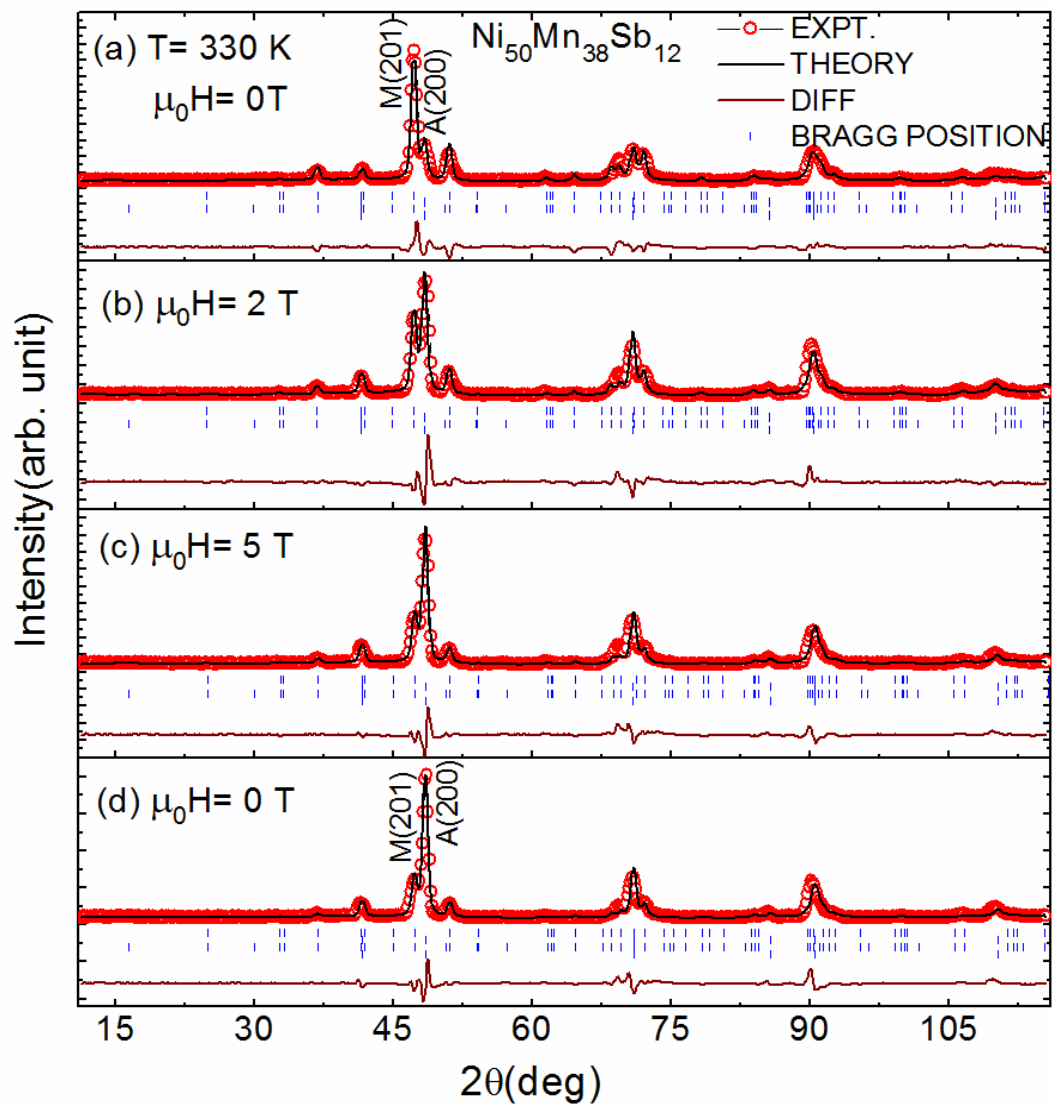

Fig. 4

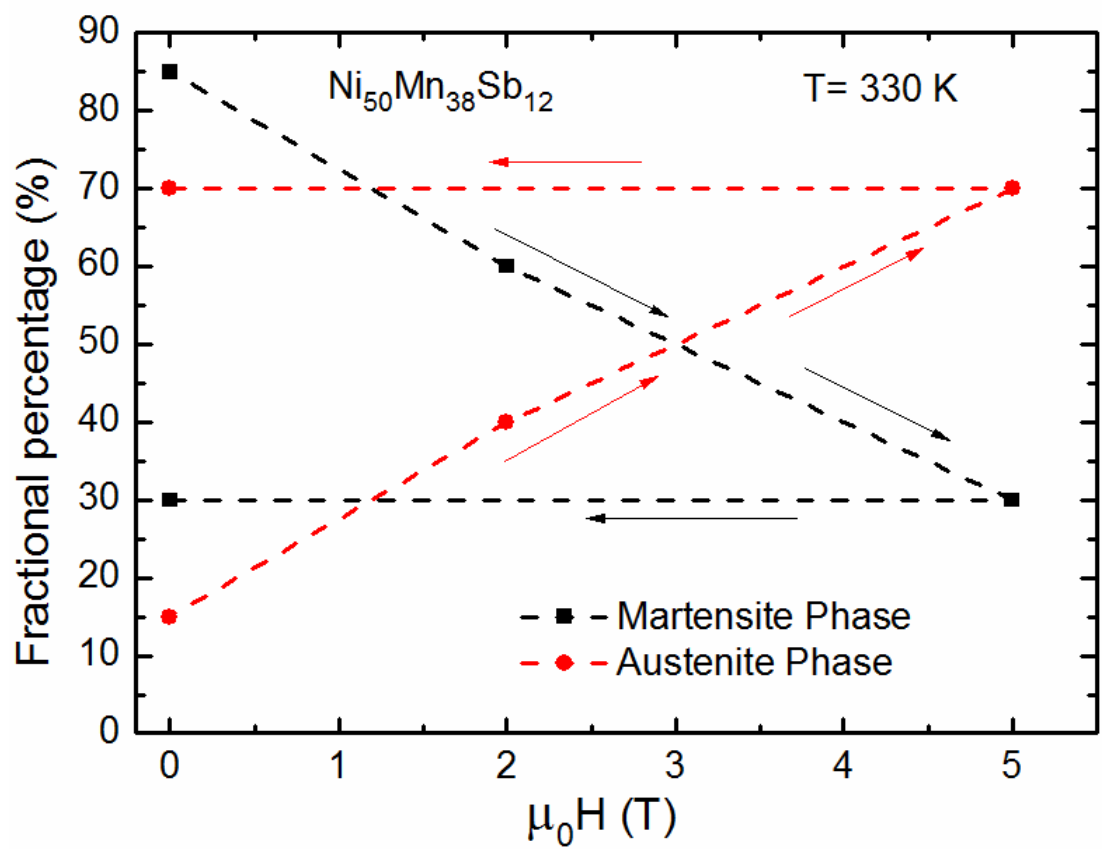